\documentclass[fleqn,twoside]{article}
\usepackage{espcrc2}

\usepackage{graphicx}
\usepackage{epsfig}

\usepackage[figuresright]{rotating}

\newcommand{\AmS}{{\protect\the\textfont2
  A\kern-.1667em\lower.5ex\hbox{M}\kern-.125emS}}

\title{ \hskip 11.5 cm {\bf \small Preprint HNINP-V-04-02} \\
Observables with $\tau$ leptons at LHC and LC \\
 structure of event records  and Monte Carlo Algorithms
}

\author{Z. W\c{a}s\address[MCSD]{Institute of Nuclear Physics\\
         Kawiory 26a, 30-055 Cracow, Poland.}%
         \thanks{This work is partly supported by
the Polish State Committee for Scientific Research 
(KBN) grant No.2P03B00122 and the European Commission 5th framework contract 
HPRN-CT-2000-00149 (`Physics at Colliders')},
}      
\begin{document}

\begin{abstract}

In the present report, let us adress the issues related to 
simulation of decays for  particle embodied in full production and decay chains
of Monte Carlo programs set-up for experiments such as at LHC  or LC.
Both technical issues related to the way how the events may be stored
in event records and issues related to physics (in particular non-factorizable
correlations of the Einstein-Rosen-Podolsky type) will be reviewed on the
basis of practical examples. 
We will limit our discussion to the case of $\tau$ lepton and $W$ boson
decays, but 
similar problems (and solutions) may arise also in case of simulation
for other intermediate states or particles. Examples related to 
construction of physics observables will be also given. In particular 
the method of measuring the CP parity properties
of the  $h-\tau\tau$ coupling at LC will be explained.  

\vspace{1pc}
\centerline{\it Presented at IX  Workshop on A C A T in Physics Research, December 1-5, 2003, KEK, Tsukuba, Japan } 

\vspace{1pc}
\end{abstract}

\maketitle

\section{Introduction}

Since many years, intensive studies are being performed to design future software architectures for
experiments on
proton proton colliders, such as the Tevatron   or the LHC \cite{LHC2}
and high energy $e^+e^-$ linear colliders such as   JLC,
NLC \cite{:2001ve} or TESLA \cite{Richard:2001qm}. 

One of the important ingredients in such designs is the data structure for storing the
Monte Carlo events. It is generally accepted that the data structures based on objects
such as particles, clusters, strings, etc.
with properties such as tracks, momenta, colour, spin, mass, etc. and on the relations
explaining the origins and descendants of the objects is the most convenient one.
This is the case at present \cite{Caso:1998tx}, and it is also envisaged for the future, see 
\cite{Boos:2001cv}. At the same time such a picture is in conflict with the basic 
principles of quantum mechanics. Einstein--Rosen--Podolsky paradox is an example of such
phenomena. A general problem is that the quantum state of a multiparticle
system cannot (at least in principle) be represented as a statistical combination of the states
defined by the products of the pure quantum states of the individual particles.
It is thus of the utmost importance to examine whether
the approximation enforced by  the data structure is purely
academic, or if it rather represents a real difficulty, which may affect the interpretation
of the  future data. In fact in some cases alternative methods can be designed and are 
in fact used as well.

It would not be a serious problem if the predictions of the Standard Model used in the 
interpretation
of the future data could be provided  by a single program, black box, without any need 
of analysing its parts. Then anything that would be measured beyond the prediction of such a
hypothetical Monte Carlo program
would be interpreted as  ``new physics''. Agreement, on the other hand, would constitute
confirmation of the Standard Model, as it is understood at present (and proper functioning 
of the detector as well). However, even in such an extreme case it is very useful,
for the purpose of experimental studies, to manipulate with the terms responsable for the
signature of the `new physics'. In this way experimental strategies can be refined,
if the new effects can be placed in well defined and phenomenologically simple modules.

Because of the complexity of the problem, Monte Carlo predictions need to be 
dealt with by programs describing: the action of the detector and of the analysis, 
on the experimental side, and various effects, such as those from hard processes, 
hadronization, decay of resonances, etc.,  on the theoretical side. 
Every part is inevitably calculated 
with some approximation, which need to be controlled.

\section{Event record and decay interface}

In the first part I will discuss solution  we used in {\tt KORALZ} \cite{Jadach:1994yv} --
the program widely used 
at LEP for the simulation  of $\tau$-lepton pair production 
and decay, including spin and QED bremssstrahlung effects. 
Even though spin effects are non-treatable
in the scheme where properties are attributed to individual particles only,
it is  the very method used there.
As described in ref. \cite{Jadach:1994yv} the algorithm 
of spin generation for  any individual event
 was consisting of the following steps:

\begin{enumerate}
\item
 An event consisting of  a pair of $\tau$ leptons, bremsstrahlung photons,
etc., was generated. 
\item
 Helicity states for both $\tau^+$ and $\tau^-$ were generated.
At this point, an   approximation with respect to quantum mechanisc was introduced.
\item
 Information on these helicty states, including the definition of quantization frames,
i.e. the relation between $\tau$'s rest frame and laboratory frame, was then  transmitted 
to {\tt TAUOLA} \cite{Jadach:1993hs,Jezabek:1991qp,Jadach:1990mz}, 
the package for the generation of $\tau$-lepton decays. 
\item
Finally  {\tt TAUOLA} performed decays of 100\% polarized $\tau$'s, and the event 
in the {\tt HEPEVT} common block was completed. 
\end{enumerate}

The solution for the spin treatment of $\tau$ leptons at LEP  
was optimal. On one side, a convenient picture
of particles with properties, origins and descendants could be used 
and, on the other, a complete  full spin solution \cite{koralb:1985,Jadach:1984ac} was available, if necessary.

Let us now turn to another example of the spin implementation algorithm.
It is taken from  ref.\cite{Pierzchala:2001gc}. The 
algorithm, essentially that of {\tt KORALZ}, was adopted to work with any Monte Carlo 
program providing the production of $\tau$-leptons. If the generated events are 
stored  in the format of a {\tt HEPEVT} common block, then  the  algorithm 
consisting  of the following basic steps can be used:
\begin{enumerate}
\item
Search for   $\tau$-leptons  in  a {\tt HEPEVT } common block 
(filled by any MC program).
\item
Check what the  origin of $\tau$--lepton is: 
$Z,\gamma,W,h,H^{\pm}$ or eventually, $2\to 2$--body process such as: 
  $e^+e^-,\; (u \bar u),\;  (d \bar d) \to \tau^+ \tau^-$.
\item
For the $2\to 2$--body process of $\tau$-pair production, it is sometimes
possible to calculate the $\tau$ polarization as a function of the invariant mass 
of the $\tau$--lepton pair and angle between the directions of  $\tau$--leptons and
incoming effective beams (in the rest frame of $\tau$-pair).
\item
If in addition to  the $\tau$-leptons,  photons
or partons (gluons, quarks, etc.) are stored in {\tt HEPEVT} common block,
one needs to define the ``effective incoming beams''.
\item
From such an information one can generate $\tau$ helicity states and define the
relation between the $\tau$ rest frame and the laboratory frame. Optionally
complete spin effects can be implemented as well, see~\cite{Was:2002gv}.
\item
The $\tau$ decay is generated with the help of  {\tt TAUOLA} and
{\tt HEPEVT } common block is appended 
with the  $\tau$'s  decay products. 

\end{enumerate}

Leading spin effects are nicely 
reproduced by the above set of programs. A more complete discussion
can be found in ref.~\cite{Pierzchala:2001gc}.

\begin{figure}[!ht]
\setlength{\unitlength}{1mm}
\begin{picture}(100,90)
\put( 0,-2){\makebox(0,0)[lb]{\epsfig{file=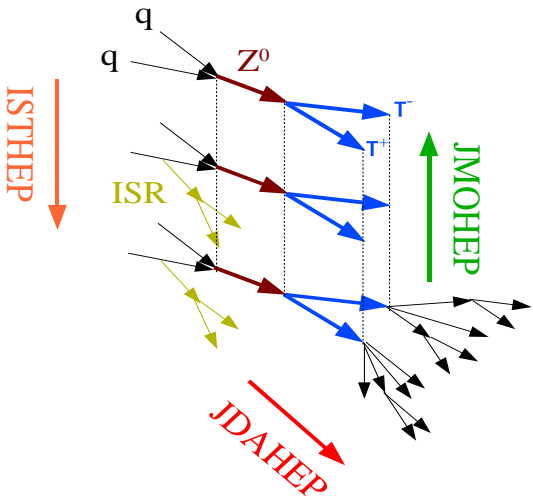,width=70mm,height=65mm}}}
\end{picture}
\caption{ \it
Typical relations in present day event record.
Daughter index {\tt JDAHEP} points upward in a tree,
however mother index  {\tt JMOHEP} may point to the previous copy of the event as well.
The status code  {\tt  ISTHEP} is ocasionally used like a pointer toward higher level
of the event record listing. As a consequence in the event record there may appear
ambiguities and/or loops for searching algorithms.
}
\label{HEPij}
\end{figure}

Let us stress, that the presented above solution, require certain minimal discipline
in a way how the event records are filled in. At present, there is a strong tendency
to store in event record, not only information on the `real particles', but also
on the results of the simulation at the parton shower level as well as of the 
hard process alone. As a consequence not only the same entries for the otherwise well 
defined
partilces like $\tau$-leptons are duplicated or even triplicated, but also 
the relation between different entries is not anymore of the tree-type and
links between particles are not reversible. The link upwards from $A$ to $B$ does
not mean that there is downward link from  $B$ to $A$. This creates multitude
of troubles for the algorithms analyzing events, see fig.~\ref{HEPij}.
However at present our solution for interfacing decay packages with the host programs
based on {\tt HEPEVT} event record as filled by PYTHIA and HERWIG seem to work in 
all cases studied by us \cite{Golonka:2003xt}. For the case of complete spin correlations, 
we found it more convenient  to abandon direct use of spin information provided by 
the host programs. Instead
we choose to calculate complete density matrix anew, from the kinematical configuration
provided by the host program.

\section{Higgs boson parity measurement}

Let us sketch the basic principles behind the proposed measurement,
in  the case, 
when simultaneosly scalar and pseudoscalar couplings
are allowed in $h\tau\tau$ vertex
\begin{equation}
\label{coupla}
h \; \bar{\tau}N(\cos\phi+i\sin\phi\gamma_{5})\tau.
\end{equation}

If non-zero CP-odd admixture to the Higgs is present, the
distribution of the Higgs production angle is modified~\cite{Abe:2001np,Kramer:1994jn,Grzadkowski:1995rx}.
We have simulated production angular distributions as in the
SM, but this assumption has no influence on the validity of the analysis.
 In order to study the
sensitivity of $h \to\tau^+\tau^-$ observables, we assume a SM production
rate inependent of the size of the CP-odd admixture.

  The production process $e^+e^- \to Z h \to \mu^+ \mu^- (q\bar q) \tau^+ \tau^- $ 
has been chosen, as an representative example, 
and simulated with the Monte Carlo program
{\tt PYTHIA 6.1} \cite{Sjostrand:2000wi}. The Higgs boson mass of 120 GeV and 
a    centre-of-mass energy of 350 GeV was chosen. 
The effects of initial state    bremsstrahlung were
included.
 For the sake of our
discussion and in all of our  samples the $\tau$ decays  have been
generated with the {\tt TAUOLA}   Monte Carlo library
\cite{Jadach:1990mz,Jezabek:1991qp,Jadach:1993hs}.  As usual, to
facilitate the interpretation of the results, bremsstrahlung effects
in decays were not taken into account. Anyway, with the help of
additional simulation, we have found this effect to be rather small.
To include the full spin effects in the    $h \to \tau^+\tau^-$,
$\tau^{\pm} \to \rho^{\pm}\bar{\nu}_{\tau}(\nu_{\tau})$,
$\rho^{\pm}\to\pi^{\pm}\pi^{0}$  decay  chain, the   interface
explained in Ref.~\cite{Was:2002gv} was used.

The Higgs boson  parity information must be extracted from   the
correlations between $\tau^{+}$ and $\tau^{-}$ spin components, which
are   further reflected in correlations between     the $\tau$ decay
products in the plane transverse to the    $\tau^{+}\tau^{-}$
axes \cite{Bower:2002zx,Desch:2003mw}. 
 To better visualize the effect, let us write  the
decay probability, using the
conventions of Ref.~\cite{Kramer:1994jn}:
\begin{equation}  
\Gamma(h_{mix}\to \tau^{+}\tau^{-}) \sim 1-s^{\tau^{+}}_{\parallel}
s^{\tau^{-}}_{\parallel}+ s^{\tau^{+}}_{\perp}
R(2\phi)~s^{\tau^{-}}_{\perp},
\label{densi}  
\end{equation}   
where $R(2\phi)$ can be understood as  an operator for the rotation by
an angle $2\phi$  around the ${\parallel}$ direction.  The
$s^{\tau^{-}}$ and $s^{\tau^{+}}$ are  the $\tau^\pm$ polarization
vectors, which are defined    in their respective rest frames. 
    The symbols ${\parallel}$/${\perp}$ denote components
parallel/transverse    to the Higgs boson momentum as seen from the
respective $\tau^\pm$  rest frames.

The method relies on measuring the acoplanarity angle
of the two planes, spanned on  $\rho^{\pm}$ decay products and defined
in the $\rho^{+}\rho^{-}$ pair rest frame.   The acoplanarity angle
$\varphi^{*}$, between the planes of the $\rho^{+}$ and $\rho^{-}$
decay products is defined.  The angle is defined first, with
the help of its cosine and two  vectors ${\bf
n}_\pm$ normal to the  planes, namely ${\bf n}_\pm= {\bf p}_{\pi^\pm}
\times  {\bf p}_{\pi^0}$, $\cos \varphi^{*} =\frac{ {\bf n}_+ \cdot
{\bf n}_-}{|{\bf n}_+| |{\bf n}_-|} $.

 To distinguish between the two cases:
$\varphi^{*}$ and  $2\pi-\varphi^{*}$ it is sufficient, for example,
to find the sign of    
$p_{\pi^-} \cdot {\bf n}_+$. When it is negative, the angle
$\varphi^{*}$ as defined above (and  in the range $0 <\varphi^{*} <
\pi$) is used.  Otherwise it is replaced by $ 2\pi - \varphi^{*}$. 

Additional selection cuts  need to be applied. 
  The events need to be  divided into two
classes, depending on the sign of  $y_{1}y_{2}$, where
\begin{equation}
y_1={E_{\pi^{+}}-E_{\pi^{0}}\over E_{\pi^{+}}+E_{\pi^{0}}}~;~~~~~
y_2={E_{\pi^{-}}-E_{\pi^{0}}\over E_{\pi^{-}}+E_{\pi^{0}}}.
\label{y1y2}
\end{equation}
The energies of $\pi^\pm,\pi^0$ are to be taken in the  respective
$\tau^\pm$ rest frames. In Refs.~\cite{Bower:2002zx,Desch:2003mw} the
methods of reconstruction of the replacement $\tau^\pm$ rest frames were
proposed with and without the help of the $\tau$ impact parameter.  We
will use these methods here as well, without any modification.

To test the feasibility of the measurement, some assumptions  about
the detector  effects had to be made, see refs.~\cite{Bower:2002zx,Desch:2003mw}
for more details. 

\subsection{  Numerical results}

\begin{figure}[!ht]
\begin{center} 
\epsfig{file=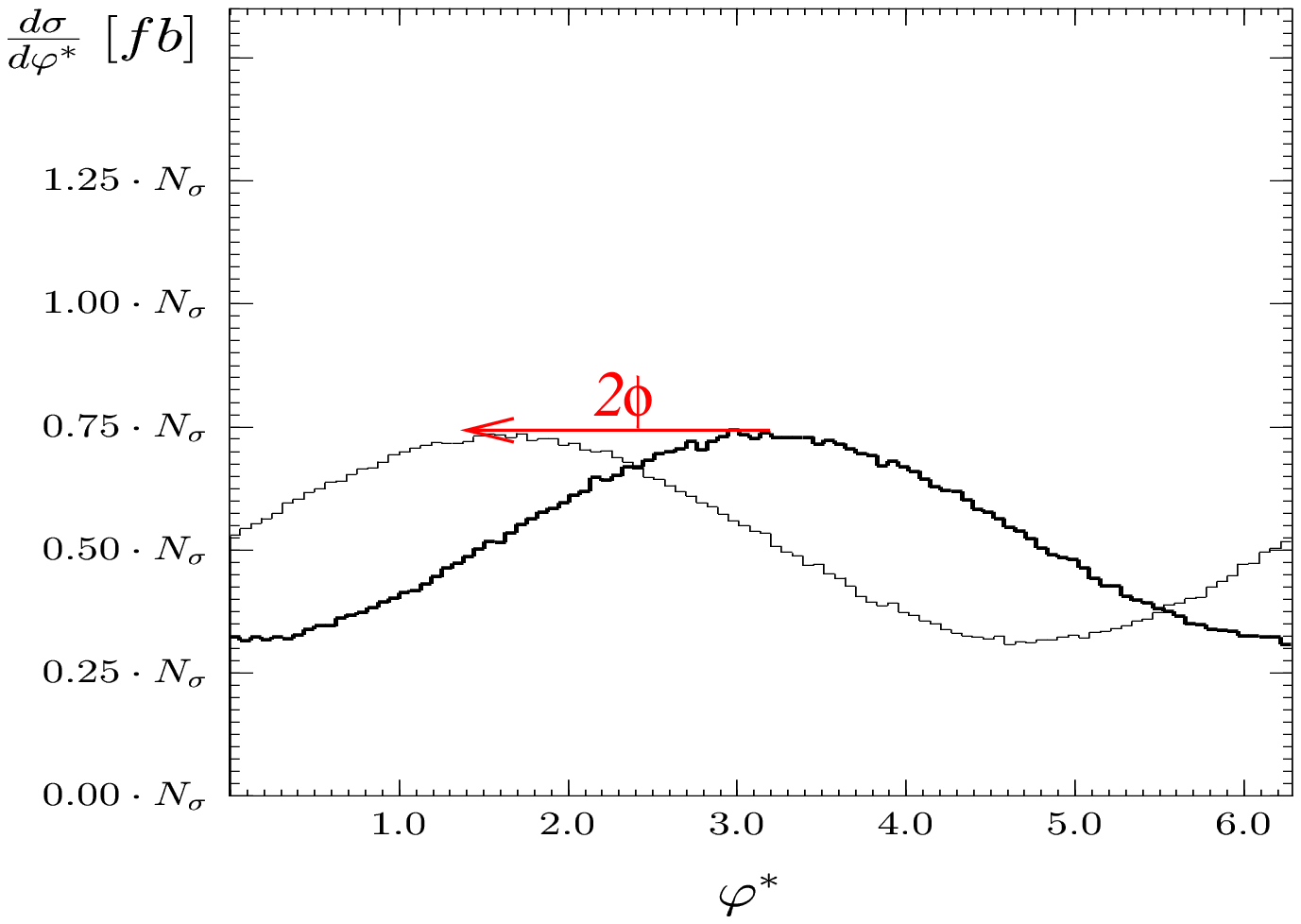,width=60mm,height=50mm}
\epsfig{file=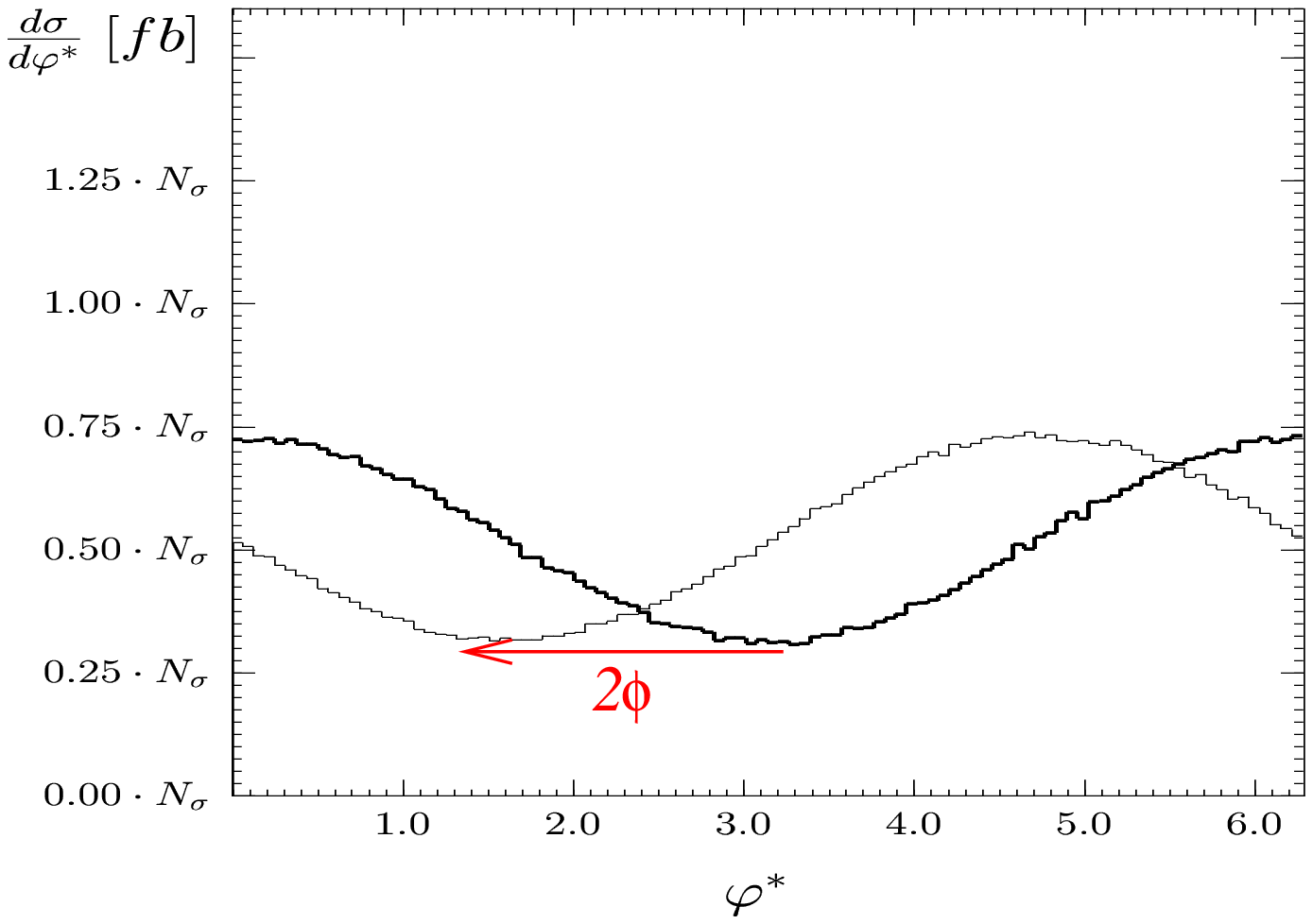,width=60mm,height=50mm}
\end{center} 
\caption  
{\it  The acoplanarity distribution  (angle $\varphi^{*}$) of the
$\rho^+ \rho^-$ decay products     in the rest frame of the $\rho^+
\rho^-$ pair. Gaussian smearing of $\pi$'s  and Higgs boson momenta,
 are included. Only events where the signs
of the  energy differences $y_1$ and $y_2$ are the same, if calculated
using the method described in Ref.~\cite{Bower:2002zx} and if
calculated  with the help of the $\tau$ impact parameter
Ref.~\cite{Desch:2003mw}, are taken.  The thick line corresponds to a
scalar Higgs boson, the thin line to a mixed one.  The left figure
contains events with $y_1 y_2 > 0$, the right one is for $y_1 y_2 <
0$. 
}
\label{rys4}  
\end{figure}

We have used the scalar--pseudoscalar mixing angle
$\phi=\frac{\pi}{4}$.
In Fig.~\ref{rys4} the acoplanarity distribution  angle $\varphi^{*}$
of the $\rho^+ \rho^-$ decay products which was defined  in the rest
frame of the reconstructed $\rho^+ \rho^-$ pair, is shown.
 The two plots represent events
selected by the differences of $\pi^\pm\pi^0$ energies, defined in
their respective $\tau^\pm$ rest frames. In the left plot, it is
required that $y_1 y_2 > 0$, whereas in the right one, events with
$y_1 y_2 < 0$ are taken. This figure quantifies the size of the parity
effect.  The size of the effect is substantially 
diminished when a detector-like set-up was included for $\tau^\pm$
rest frames reconstruction, in exactly
the same proportion as in Ref.~\cite{Bower:2002zx}, nonetheless 
parity effect remain visible.

The fitting  procedure was repeated 400
times with acoplanarity distributions extracted from independent
samples of 1~ab$^{-1}$ luminosity each, with a nominal value of $\phi
= \pi/4$. A precision on $\phi$ from such a pseudo-experiment of
approximately 6$^{\circ}$ can be anticipated.

\section{  Summary}

The combination of generators for production and decay of intermediate states, 
require careful treatment of the spin degrees of freedom. In some cases one can restrict
spin states to pure helicities; then generation of intermediate states for individual
particles can be performed first, and decays of each individual particle
can be performed later. 
The general case, when full quantum mechanical spin correlations are included was
also discussed. Technical constraints for the solution based on kinematical 
information provided by the production programme  to be used by the decay routines,
were presented. In this context gramatic rules on how event records are filled in were discussed as well.

Finally discussion of observable for the Higgs boson parity measurement at LC,
based on such a technical solution was presented in detail, as an example. 
It was shown, on the basis of 
careful Monte Carlo simulation of both theoretical and detector effects that with the
typical parameters of the future detector and Linear Collider set-up the hypotesis
of the admixture of pseudoscalar coupling to the {\it otherwise} Standard Model 120 GeV Higgs boson
can be measured up to 6$^o$ error on the mixing angle.

\section{ Acknowledgements}

I am grateful to co-authors: 
G. Bower, C. Biscarat, K. Desch, P. Golonka, A. Imhof,
B. Kersevan, T. Pierzcha\l a, E. Richter-Was
and M. Worek 
of the papers and related activities which lead to the presented talk.

\providecommand{\href}[2]{#2}
\end{document}